# Integration of the Noncollinear Antiferromagnetic Metal Mn$_3$Sn onto Ferroelectric Oxides for Electric-Field Control

Xiaoning Wang, Zexin Feng, Peixin Qin, Han Yan, Xiaorong Zhou, Huixin Guo, Zhaoguogang Leng, Weiqi Chen, Qiannan Jia, Zexiang Hu, Haojiang Wu, Xin Zhang, Chengbao Jiang, Zhiqi Liu*

X. Wang, Z. Feng, P. Qin, H. Yan, X, Zhou, H. Guo, Z. Leng, W. Chen, Q. Jia, Z. Hu, H. Wu, X. Zhang, Prof. C. Jiang, Prof. Z. Liu
School of Materials Science and Engineering, Beihang University, Beijing 100191, China

*Email: zhiqi@buaa.edu.cn

Keywords: Mn$_3$Sn, Weyl antiferromagnetic metal, electric-field control, multiferroic heterostructures, antiferromagnetic piezospintronics

**Abstract:** Non-collinear antiferromagnetic materials have received dramatically increasing attention in the field of spintronics as their exotic topological features such as the Berry-curvature-induced anomalous Hall effect and possible magnetic Weyl states could be utilized in future topological antiferromagnetic spintronic devices. In this work, we report the successful integration of the antiferromagnetic metal Mn$_3$Sn thin films onto ferroelectric oxide PMN-PT. By optimizing growth, we realized the large anomalous Hall effect with small switching magnetic fields of several tens mT fully comparable to those of bulk Mn$_3$Sn single crystals, anisotropic magnetoresistance and negative parallel magnetoresistance in Mn$_3$Sn thin films with antiferromagnetic order, which are similar to the signatures of the Weyl state in bulk Mn$_3$Sn single crystals. More importantly, we found that the anomalous Hall effect in antiferromagnetic Mn$_3$Sn thin films can be manipulated by electric fields applied onto the ferroelectric materials, thus demonstrating the feasibility of Mn$_3$Sn-based topological spintronic devices operated in an ultralow power manner.

# 1. Introduction

Since the theoretical prediction and subsequent experimental realization of the anomalous Hall effect (AHE) in non-collinear antiferromagnets [1-6], Mn$_3$Sn has been a hot spot in antiferromagnetic (AFM) spintronics. For example, the magnetic Weyl state has been theoretically predicted [7,8] and the evidence of this exotic state has been found for Mn$_3$Sn [9]. More recently, Mn$_3$Sn has been used in AFM spintronic devices, where the manipulation of the non-collinear spins enables the magnetic spin Hall and magnetic inverse spin Hall effects [10].

The advantages of using AFM materials in spintronic devices are obvious in terms of the spin dynamics and robustness over external magnetic perturbations. Thus, AFM spintronics has developed fast in recent years [11-24]. It could revolutionarily replace the current information storage technique for picosecond information writing [25-27], higher integration density, and magnetic-field insensitive information storage [24,28].

To use an AFM material for memory device applications, one must manage to control its resistance state. Regarding this aspect, either the longitudinal resistance can be modified via the anisotropic magnetoresistance (AMR) effect [29-31] by spin-orbit torque [30] or piezoelectric strain [28], or the transverse Hall resistance can be reversed by magnetic fields [3,10] or piezoelectric strain [6]. Among these approaches, both the spin-orbit torque and magnetic fields rely on electrical currents which generate considerable Joule heating during information writing and accordingly need high energy consumption. In contrast, the piezoelectric strain approach could cost ultralow power owing to the suppression of Joule heating as a result of the highly insulating nature of ferroelectric materials [23,24]. Therefore, it is highly desired that the Weyl antiferromagnet Mn$_3$Sn could be integrated onto ferroelectric materials to realize electric-field control in topological AFM spintronic devices. In this work, we report the success of the endeavor on this important perspective.

## 2. Experimental methods

Mn$_3$Sn films were grown on (001)-oriented MgO, 0.7PbMg$_{1/3}$Nb$_{2/3}$O$_3$–0.3PbTiO$_3$ (PMN-PT), 100-nm-thick LaAlO$_3$-buffered PMN-PT substrates from a polycrstalline Mn$_3$Sn target by a d.c. sputtering system with a base pressure of 7.5×10$^{-9}$ Torr. The growth temperature was 150 ˚C. The sputtering power and the Ar pressure during deposition were 60 W and 3 mTorr, respectively. The growth rate was ~0.28 Å/s as determined by transmission electron microscopy measurements. Co$_{90}$Fe$_{10}$ (CoFe) thin films were grown by the d.c. sputtering system as well. The sputtering power and the Ar pressure were 90 W and 3 mTorr, respectively. The growth rate was ~0.11 Å/s. Pt thin films were sputtered by 30 W at an Ar pressure of 3 mTorr. Its growth rate was 0.5 Å/s. The 100-nm-thick LaAlO$_3$ buffer layers on PMN-PT substrates were fabricated by pulsed laser deposition with a laser fluence of ~1.6 J/cm$^2$, an oxygen pressure of 10$^{-2}$ Torr and a repetition rate of 10 Hz at 800 ˚C. The focused laser spot size was 1×3 mm$^2$ and the target-substrate distance was 60 mm. The deposition rate of LaAlO$_3$ is ~0.11 Å/pulse.

A Quantum Design VersaLab system was used for conducting electrical and magnetic measurements. The measuring current was 1 mA supplied from a Keithley 2400 sourcemeter. A Keithley 2182A nanovolt meter was used to collect voltage signals for both longitudinal and transverse resistance measurements. The gate electric field was supplied by another Keithley 2400 sourcemeter.

## 3. Results and discussion

Firstly, by varying the growth temperature of Mn$_3$Sn thin films onto (001)-oriented MgO single-crystal substrates from room temperature to 750 ˚C, we identified that the optimal growth temperature for achieving the most remarkable AHE and the lowest magnetic field for switching the anomalous Hall resistance is 150 ˚C. As shown in Fig. 1a, a 50-nm-thick Mn$_3$Sn/MgO heterostructure fabricated at 150 ˚C exhibit a rather sharp interface between the Mn$_3$Sn film and the MgO substrate. The Mn$_3$Sn film is reasonably flat and continuous. The zoom-in image of an interfacial region (Fig. 1b) reveals the textured polycrystalline feature of the Mn$_3$Sn film with epitaxy in the first several atomic layers. X-ray diffraction measurements reveal a preferred (002) orientation with an average grain size of ~16 nm. Compared with the crystallization temperature of oxides that is typically ~450 ˚C [32], the intermetallic Mn$_3$Sn starts to crystallize at a much lower temperature.

Diamagnetic oxide substrates could contain ferromagnetic impurities and thus exhibit ferromagnetic-like signals [33], which could be misleading especially for magnetic measurements of thin films with small magnetic signals. To check this out, we examined our MgO substrates by magnetic measurements and it was found that they show linear diamagnetic signals, which ensures the reliable analysis of ferromagnetic signals of Mn$_3$Sn thin films grown on them. The room-temperature magnetic signal of a Mn$_3$Sn/MgO heterostructure consists a dominant diamagnetic moment from the MgO substrate and a weak ferromagnetic moment of the Mn$_3$Sn film (Fig. 2a). After subtraction of the substrate signal, a saturation magnetization of ~5.1 emu/cc (Fig. 2b) is obtained, which is typical for the weak ferromagnetism in canted antiferromagnets such as in BiFeO$_3$ [34]. This magnetization corresponds to an average magnetic moment of ~12 m$\mu_B$/Mn. It is larger than the bulk value ~2 m$\mu_B$/Mn but rather comparable with the moment in Mn$_3$Sn films deposited on Si substrates [35,36].

To explore the intrinsic AFM order of the deposited Mn$_3$Sn thin films, we investigated their exchange coupling with a soft ferromagnetic (FM) Co$_{90}$Fe$_{10}$ (CoFe) layer [28,37]. Fig. 3a plots the magnetic-field-dependent magnetization (*M-H*) of a 5-nm-thick CoFe layer grown on a MgO single-crystal substrate capped by a 2-nm-thick Pt layer at different temperatures. At room temperature, its coercivity field $\mu_0H_c$ is ~4.6 mT. It greatly increases with lowering temperature and reaches ~19.3 mT at 50 K. Its saturation magnetization is ~1200 emu/cc at room temperature and is consistent with Ingvarssona *et al.*'s results for CoFe thin films [38].

In sharp contrast, a CoFe film grown on a 50-nm-thick $Mn_3Sn$ film exhibits significantly large coercivity fields (Fig. 3b). For example, $H_c$ is ~6.6 mT and ~51.3 mT at 300 and 50 K, respectively, which is enhanced by ~44% at 300 K and ~166% at 50 K compared with $H_c$ of a single CoFe layer. For all the temperatures ranging from 50 to 300 K, the enhancement of $H_c$ can be clearly seen as summarized in Fig. 3c. In exchange-coupled AFM/FM bilayer systems, either the exchange bias or the enhancement of the coercivity field of the FM layer or both could appear as result of the exchange coupling [39,40]. Thus, such a giant enhancement of the coercivity field proves the very strong exchange coupling between $Mn_3Sn$ and CoFe, which also evidences the antiferromagnetic order in $Mn_3Sn$ thin films.

The Hall effect measurements were subsequently performed. It was found that the AHE exists at all temperatures (Fig. 4). Below 200 K, the AHE exhibits obvious hysteresis loop as first discovered by Nakatsuji *et al*. [3] At 50 K and 1 T, the magnitude of anomalous Hall resistance $R_{Hall}$ is ~0.37 Ω, corresponding to a Hall resistivity of ~1.85 μΩ·cm, which is on the same order with that of bulk $Mn_3Sn$ [3]. Remarkably, the magnetic field for switching the anomalous Hall resistance in our $Mn_3Sn$ films is ~90, ~80, ~65 and ~20 mT for 50, 75, 100 and 150 K, respectively, which is rather comparable with that of bulk $Mn_3Sn$ [3]. It has been a challenge to realize low switching fields of the AHE in $Mn_3Sn$ thin films as it is more than 1 T in thin films [35,36,41], which is 20 times larger than that of bulk $Mn_3Sn$ (few ten mT) and prevents $Mn_3Sn$ thin films for high-density spintronic device applications. For example, the $Mn_3Sn$ films annealed at 500 °C by Higo *et al*. [35] and Ikeda *et al*. [36] show a magnetic switching field of ~1 T; You *et al*. [41] grew quasi epitaxial $Mn_3Sn$ films on MgO at 420 °C and the resulting magnetic switching fields are large than 1 T. The large switching fields in high-temperature grown $Mn_3Sn$ thin films could be related to antisite defects such as them in high-temperature grown antiferromagnetic MnPt films [37]. That is because intermetallic alloys possess largely different surface energies compared with oxide substrate materials and this difference could be even large at high temperatures, consequently leading to the balling issue and defects due to the wetting problem [23].

As a result, the current $Mn_3Sn$-based spintronic devices have been only demonstrated based on thinned $Mn_3Sn$ bulk crystals [10]. The small switching magnetic fields realized in our $Mn_3Sn$ thin films would largely facilitate magnetic-field control of the non-collinear AFM spin structure and anomalous Hall resistance in $Mn_3Sn$-based spintronic devices.

Following the same thin film growth procedure, we started to grow $Mn_3Sn$ thin films on (001)-oriented $0.7PbMg_{1/3}Nb_{2/3}O_3$–$0.3PbTiO_3$ (PMN-PT) single-crystal substrates. In general, the integration of intermatallic alloys on ferroelectric PMN-PT is rather challenging and needs lots of practice [42]. The reason is that Pb tends to be volatile at high temperatures and consequenlty elements of intermetallic alloys would diffuse into Pb vacacies, leading to oxidation and secondary alloys. However, the optimized growth temperature of 150 ˚C for $Mn_3Sn$ thin film growth is relatively low and thus may allow transplating high crystal quanlity and exotic physical properties of $Mn_3Sn$ films onto PMN-PT.

As shown in Fig. 5, the magnetic signal of a 50-nm-thick $Mn_3Sn$/PMN-PT heterostructure is akin to that of the $Mn_3Sn$/MgO heterostructure (Fig. 2). The saturation magnetization of the weak ferromagnetism in the $Mn_3Sn$ film grown on PMN-PT is ~4 emu/cc, which is slightly smaller than that of $Mn_3Sn$/MgO heterostructures (Fig. 2b). This implies the suppression of the AFM spin canting. In addition, $Mn_3Sn$ thin films deposited onto PMN-PT substrates exhibit a saturation field of ~0.25 T, which is larger than that of a $Mn_3Sn$/MgO heterostructure, ~0.15 T (Fig. 2b). Similar to the $Mn_3Sn$/MgO heterostructure, the AHE in the $Mn_3Sn$/PMN-PT heterostructure is visible for all the temperatures and the feature of the small switching magnetic field remains, which is ~128, ~110, ~96, ~56, ~50, and ~50 mT for 50, 75, 100, 150, 200 and 300 K, respectively. (Fig. 6).

Weyls fermions are a new class of topological state of matter. In a Weyl fermion system, Weyl points, originating from the linear crossing of non-degenerate bands near the Fermi level, typically appear in pair with opposite chirality and are connected by Fermi arcs in the momentum space [43]. From the angle of magnetotransport, the exotic Weyl state results in AMR and negative magnetoresistance when the measuring current is parallel to the magnetic field, which is believed as the signature of chiral anomaly [44]. In bulk $Mn_3Sn$, the existence of Weyl points is evidenced by angle-resolved photoemission spectroscopy measurements [9]. Meanwhile, both the AMR and negative parallel magnetoresistance are shown to exist.

Armed with the above thought, we examined the magnetoresistance of $Mn_3Sn$ films grown on PMN-PT. As plotted in Fig. 7a, although the magnitude of the parallel negative magnetoresistance is very small, on the order of ~0.01%, it can be clearly seen for almost all the temperatures up to 3 T. In addition, as shown in the polar figure of Fig. 7b, the AMR with twofold symmetry over 360° is obvious for all the temperatures ranging from 50 to 300 K. These data are similar to the signatures of the exotic Weyl state in bulk $Mn_3Sn$ single crystals.

After the successful integration of $Mn_3Sn$ thin films onto PMN-PT substrates, we then used electric fields perpendicularly applied across the ferroelectric substrate to modulate electrical properties of $Mn_3Sn$ thin films. To our disappointment, PMN-PT substrates become very fragile after the deposition of $Mn_3Sn$ thin films and tend to crack easily especially when a negative gate electric field $E_G$ (schematized in Fig. 8a) is applied, leading to a colossal resistance change in the $Mn_3Sn$ films at room temperature as plotted in Fig. 8b. Cracks-induced colossal resistance change has been systematically studied by our recent work and it was found that cracks are mainly induced by the internal stress accumulated at domain boundaries between freely switchable ferroelectric domains and other domains pinned by surface defects [45]. It is possible that the deposition of $Mn_3Sn$ on PMN-PT induces more defects on the surfaces of PMN-PT single crystals, which behave as pinning points and are responsible for poor endurance upon external periodic gate electric fields.

To solve this problem, we tentatively deposited a 100-nm-thick LaAlO$_3$ (LAO) buffer layer onto a PMN-PT substrate at a high oxygen pressure of 10$^{-2}$ Torr (Fig. 9a). As LAO is a chemically stable and highly insulating oxide with a large band of ~5.6 eV [46], the insert of a LAO layer could prevent any interfacial chemical reaction between PMN-PT and chemically active Mn$_3$Sn. It turns out that such a buffer layer enhances the mechanical properties of PMN-PT significantly and greatly prevents PMN-PT from cracking.

As shown in Fig. 9b, the AHE is available in a Mn$_3$Sn film grown on a LAO-buffered PMN-PT substrate as well. Then we turned to examine the effect of a gate electric field $E_G$ (Fig. 10a) on the AHE of Mn$_3$Sn that is induced by the topological features of Bloch bands and the resulting non-vanishing Berry curvature. At 150 K, an $E_G$ of -3.6 kV/cm, which was applied at room temperature and kept onto the heterostructure during cooling down to 150 K, noticeably enhances the anomalous Hall resistance (Fig. 10b). At zero magnetic field, $R_{Hall}$ is increased from ~4.1 at $E_G$ = 0 kV/cm to ~11.9 mΩ at $E_G$ = -3.6 kV/cm, corresponding to a ~190% anomalous Hall resistance enhancement. This signifies the application potential of the anomalous Hall resistance for information encoding operated by electric fields at zero magnetic field. A similar trend of the enhancement of $R_{Hall}$ by a gate electric field is found at 200 K as well (Fig. 10c).

As LAO has a much lower dielectric constant (~25) compared with ferroelectric PMN-PT, the addition of the LAO layer between PMN-PT and intermetallic Mn$_3$Sn would largely reduced the electrostatic effect. On the hand, the carrier density of Mn$_3$Sn is close to that of usual metals, ~6.8×10$^{22}$ /cm$^3$, and thus the Thomas-Fermi screening length is very short, on the order of angstrom [47], which is much less than our film thickness 50 nm. Therefore, the predominant mechanism for the electric-field-modulated AHE should be piezoelectric strain [28].

Generally, the anomalous Hall effect in Mn$_3$Sn originates from the triangular antiferromagnetic spin structure induced non-vanishing Berry curvature as theoretically predicted [1] and experimentally demonstrated [3] However, to enable the switching of the triangular spin structure and the resulting anomalous Hall resistance by magnetic fields, a subtle canted magnetic moment is the key, which responds to an external magnetic field and help coherently switch the order of the triangular spin order and the anomalous Hall effect. The piezoelectric strain exerted on Mn$_3$Sn thin films can modify the magnetic anisotropy of antiferromagnetic spins. The strain-modulated anisotropy has been demonstrated for noncollinear antiferromagnet Mn$_3$Pt [6], collinear antiferromagnet MnPt [28], and other Mn-based collinear antiferromagnets [48]. More relevantly, Lukashev *et al*. investigated the piezomagnetic effect in Mn-based noncollinear antiferromagnet and found that biaxial compressive piezoelectric strain can rotate the triangular spin structure so that the resulting Berry-curvature-related Hall vector may exhibit a larger projection along the out-of-plane direction[49], which thus enhances the AHE in antiferromagnets.

Although chiral spintronics involving non-collinear spins have been mainly achieved in intermetallic magnetic alloy systems, it could be fully feasible in other material systems such as strongly correlated oxides. For example, the topological Hall effect has been recently observed [50] in single-layer atomically thin SrRuO$_3$ thin films originating from the tilting of oxygen octahedra by epitaxial strain [51,52]. From this perspective, the interesting chiral spintronic phenomena including the non-vanishing Berry curvature, the spin Hall effect without spin-orbit coupling [53], the topological Hall effect and the skyrmion phase could be realized in other magnetic oxides such as LaMnO$_3$ [54,55], LaNiO$_3$ [56], and LaCoO$_3$ [57] by strain engineering of oxygen octahedra.

## 4. Conclusions

In summary, we have successfully integrated the noncollinear antiferromagnet $Mn_3Sn$ onto ferroelectric PMN-PT in term of thin films, which is essential for utilizing this exotic material for advanced topological spintronic devices. Especially, the low switching fields of few ten mT achieved in our $Mn_3Sn$ thin films is comparable to that of bulk $Mn_3Sn$ and would largely facilitate the magnetic-field switching of the non-collinear AFM spin structure for device applications. The magnetotransport study reveals the similarity between our low-temperature-fabricated $Mn_3Sn$ thin films and the Weyl antiferromagnetic bulk $Mn_3Sn$ single crystals. More importantly, we have further demonstrated a proof-of-concept of an electric-field-controlled topological antiferromagnetic spintronic device [20,23] based on a $Mn_3Sn$ thin film that has been highly desired from the perspective of energy consumption.

## Acknowledgements
Zhiqi Liu acknowledges financial support from the National Natural Science Foundation of China (NSFC; grant numbers 51822101, 51861135104, 51771009 & 11704018).

**References:**

1. H. Chen, Q. Niu, A. H. MacDonald, Anomalous Hall effect arising from noncollinear antiferromagnetism, Phys. Rev. Lett. 112 (2014) 017205.
2. J. Kübler, C. Felser, Non-collinear antiferromagnets and the anomalous Hall effect, Europhys. Lett. 108 (2014) 67001.
3. S. Nakatsuji, N. Kiyohara, T. Higo, Large anomalous Hall effect in a non-collinear antiferromagnet at room temperature, Nature 527 (2015) 212.
4. N. Kiyohara, T. Tomita, S. Nakatsuji, Giant anomalous Hall effect in the chiral antiferromagnet $Mn_3Ge$, Phys. Rev. Appl. 5 (2016) 064009.
5. A. K. Nayak, J. E. Fischer, Y. Sun, B. Yan, J. Karel, A. C. Komarek, C. Shekhar, N. Kumar, W. Schnelle, J. Kübler, C. Felser, S. S. P. Parkin, Large anomalous Hall effect driven by a nonvanishing Berry curvature in the noncolinear antiferromagnet $Mn_3Ge$, Sci. Adv. 2 (2016) e1501870.
6. Z. Q. Liu, H. Chen, J. M. Wang, J. H. Liu, K. Wang, Z. X. Feng, H. Yan, X. R. Wang, C. B. Jiang, J. M. D. Coey, A. H. MacDonald, Electrical switching of the topological anomalous Hall effect in a non-collinear antiferromagnet above room temperature, Nat. Electron. 1 (2018) 172.
7. J. Liu, L. Balents, Anomalous Hall effect and topological defects in antiferromagnetic Weyl semimetals: $Mn_3Sn/Ge$, Phys. Rev. Lett. 119 (2017) 087202.
8. J. Kübler, C. Felser, Weyl fermions in antiferromagnetic $Mn_3Sn$ and $Mn_3Ge$, Europhys. Lett. 120 (2017) 47002.
9. K. Kuroda, T. Tomita, M. –T. Suzuki, C. Bareille, A. A. Nugroho, P. Goswami, M. Ochi, M. Ikhlas, M. Nakayama, S. Akebi, R. Noguchi, R. Ishii, N. Inami, K. Ono, H. Kumigashira, A. Varykhalov, T. Muro, T. Koretsune, R. Arita, S. Shin, T. Kondo，S. Nakatsuji, Evidence for magnetic Weyl fermions in a correlated metal, Nat. Mater. 16 (2017) 1090.
10. M. Kimata, H. Chen, K. Kondou, S. Sugimoto, P. K. Muduli, M. Ikhlas, Y. Omori, T. Tomita, A. H. MacDonald, S. Nakatsuji, Y. Otani, Magnetic and magnetic inverse spin Hall effects in a non-collinear antiferromagnet, Nature 565 (2019) 627.
11. A. H. MacDonald, M. Tsoi, Antiferromagnetic metal spintronics, Phil. Trans. R. Soc. A 369 (2011) 3098.
12. E. V. Gomonay, V. M. Loktev, Spintronics of antiferromagnetic systems, Low Temp. Phys. 40 (2014) 17.

13. T. Jungwirth, X. Marti, P. Wadley, J. Wunderlich, Antiferromagnetic spintronics, Nat. Nanotechnol. 11 (2016) 231.
14. V. Baltz, A. Manchon, M. Tsoi, T. Moriyama, T. Ono, Y. Tserkovnyak, Antiferromagnetic spintronics, Rev. Mod. Phys. 90 (2018) 015005.
15. O. Gomonay, T. Jungwirth, J. Sinova, Concepts of antiferromagnetic spintronics, Phys. Status Solidi RRL 11 (2017) 1700022.
16. O. Gomonay, V. Baltz, A. Brataas, Y. Tserkovnyak, Antiferromagnetic spin textures and dynamics, Nat. Phys. 14 (2018) 213.
17. R. A. Duine, K. J. Lee, S. S. P. Parkin, M. D. Stiles, Synthetic antiferromagnetic spintronics, Nat. Phys. 14 (2018) 217.
18. J. Železný, P. Wadley, K. Olejník, A. Hoffmann, H. Ohno, Spin transport and spin torque in antiferromagnetic devices, Nat. Phys. 14 (2018) 220.
19. T. Jungwirth, J. Sinova, A. Manchon, X. Marti, J. Wunderlich, C. Felser, The multiple directions of antiferromagnetic spintronics, Nat. Phys. 14 (2018) 200.
20. L. Šmejkal, Y. Mokrousov, B. Yan, A. H. MacDonald, Topological antiferromagnetic spintronics, Nat. Phys. 14 (2018) 242.
21. P. Němec, M. Fiebig, T. Kampfrath, A. V. Kimel, Antiferromagnetic opto-spintronics, Nat. Phys. 14 (2018) 229.
22. M. B. Jungfleisch, W. Zhang, A. Hoffmann, Perspectives of antiferromagnetic spintronics, Phys. Lett. A 382 (2018) 865.
23. Z. Feng, H. Yan, Z. Liu, Electric-field control of magnetic order: from FeRh to topological antiferromagnetic spintronics, Adv. Electron. Mater. 5 (2018) 1800466.
24. Z. Liu, Z. Feng, H. Yan, X. Wang, X. Zhou, P. Qin, H. Guo, R. Yu, C. Jiang, Antiferromagnetic Piezospintronics, Adv. Electron. Mater. 5 (2019) 1900176.
25. A. V. Kimel, A. Kirilyuk, A. Tsvetkov, R. V. Pisarev, T. Rasing, Laser-induced ultrafast spin reorientation in the antiferromagnet $TmFeO_3$, Nature, 429 (2004) 850.
26. K. Olejník, T. Seifert, Z. Kašpar, V. Novák, P. Wadley, R. P. Campion, M. Baumgartner, P. Gambardella, P. Němec, J. Wunderlich, J. Sinova, P. Kužel, M. Müller, T. Kampfrath, T. Jungwirth, Terahertz electrical writing speed in an antiferromagnetic memory, Sci. Adv. 4 (2018) eaar3566.
27. J. Tang, Y. Ke, W. He, X. Zhang, N. Li, Y. Zhang, Y. Li, Z. Cheng, Ultrafast photoinduced multimode antiferromagnetic spin dynamics in exchange-coupled Fe/$RFeO_3$ (R = Er or Dy) heterostructures, Adv. Mater. 30 (2018) 1706439.

28. H. Yan, Z. Feng, S. Shang, X. Wang, Z. Hu, J. Wang, Z. Zhu, H. Wang, Z. Chen, H. Hua, W. Lu, J. Wang, P. Qin, H. Guo, X. Zhou, Z. Leng, Z. Liu, C. Jiang, M. Coey, Z. Liu, A piezoelectric, strain-controlled antiferromagnetic memory insensitive to magnetic fields, Nat. Nanotechnol. 14 (2019) 131.

29. X. Marti, I. Fina, C. Frontera, J. Liu, P. Wadley, Q. He, R. J. Paull, J. D. Clarkson, J. Kudrnovský, I. Turek, J. Kuneš, D. Yi, J. H. Chu, C. T. Nelson, L. You, E. Arenholz, S. Salahuddin, J. Fontcuberta, T. Jungwirth, R. Ramesh, Room-temperature antiferromagnetic memory resistor, Nat. Mater. 13 (2014) 367.

30. P. Wadley, B. Howells, J. Železný, C. Andrews, V. Hills, R. P. Campion, V. Novák, K. Olejník, F. Maccherozzi, S. S. Dhesi, S. Y. Martin, T. Wagner, J. Wunderlich, F. Freimuth, Y. Mokrousov, J. Kuneš, J. S. Chauhan, M. J. Grzybowski, A. W. Rushforth, K. W. Edmonds, B. L. Gallagher, T. Jungwirth, Electrical switching of an antiferromagnet, Science 351 (2016) 587.

31. C. Lu, B. Gao, H. Wang, W. Wang, S. Yuan, S. Dong, J. -M. Liu, Anisotropic magnetoresistance in an antiferromagnetic semiconductor, Adv. Funct. Mater. 28 (2018) 1706589.

32. Z. Q. Liu, W. Lu, S. W. Zeng, J. W. Deng, Z. Huang, C. J. Li, M. Motapothula, W. M. Lü, L. Sun, K. Han, J. Q. Zhong, P. Yang, N. N. Bao, W. Chen, J. S. Chen, Y. P. Feng, J. M. D. Coey, T. Venkatesan, Ariando, Bandgap control of the oxygen-vacancy-induced two-dimensional electron gas in $SrTiO_3$, Adv. Mater. Interfaces 1 (2014) 1400155.

33. M. Khalid, A. Setzer, M. Ziese, P. Esquinazi, D. Spemann, A. Pöppl, and E. Goering, Ubiquity of ferromagnetic signals in common diamagnetic oxide crystals, Phys. Rev. B 81 (2010) 214414.

34. Z. Chen, J. Liu, Y. Qi, D. Chen, S. –L, Hsu, A. R. Damodaran, X. He, A. T. N'Diaye, A. Rockett, L. W. Martin, 180° ferroelectric stripe nanodomains in $BiFeO_3$ thin films, Nano. Lett. 15 (2015) 6506.

35. T. Higo, D. Qu, Y. Li, C. L. Chien, Y. Otani, S. Nakatsuji, Anomalous Hall effect in thin films of the Weyl antiferromagnet $Mn_3Sn$, Appl. Phys. Lett. 113 (2018) 202402.

36. T. Ikeda, T. Ikeda, M. Tsunoda, M. Oogane, S. Oh, T. Morita, Y. Ando, Anomalous Hall effect in polycrystalline $Mn_3Sn$ thin films, Appl. Phys. Lett. 113 (2018) 222405.

37. Z. Liu, M. D. Biegalski, S. L. Hsu, S. Shang, C. Marker, J. Liu, L. Li, L. Fan, T. L. Meyer, A. T. Wong, J. A. Nichols, D. Chen, L. You, Z. Chen, K. Wang, K. Wang, T. Z. Ward, Z. Gai, H. N. Lee, A. S. Sefat, V. Lauter, Z. K. Liu, H. M. Christen, Epitaxial growth of intermetallic MnPt films on oxides and large exchange bias, Adv. Mater. 28 (2016) 118.

38. S. Ingvarssona, G. Xiao, S.S.P. Parkin, W.J. Gallagher, Thickness-dependent magnetic properties of $Ni_{81}Fe_{19}$, $Co_{90}Fe_{10}$ and $Ni_{65}Fe_{15}Co_{20}$ thin films, J. Magn. Magn. Mater. 251 (2002) 202.

39. C. Leighton, J. Nogués, B. J. Jönsson-Åkerman, I. K. Schuller, Coercivity enhancement in exchange biased systems driven by interfacial magnetic frustration, Phys. Rev. Lett. 84 (2000) 3466.

40. C. A. F. Vaz, Electric field control of magnetism in multiferroic heterostructures, J. Phys.: Condens. Matter 24 (2012) 333201.

41. Y. You, X. Chen, X. Zhou, Y. Gu, R. Zhang, F. Pan, C. Song, Anomalous Hall effect–like behavior with in-plane magnetic field in noncollinear antiferromagnetic $Mn_3Sn$ films, Adv. Eletron. Mater. 5 (2019) 1800818.

42. Y. Lee, Z. Q. Liu, J. T. Heron, J. D. Clarkson, J. Hong, C. Ko, M. D. Biegalski, U. Aschauer, S. L. Hsu, M. E. Nowakowski, J. Wu, H. M. Christen, S. Salahuddin, J. B. Bokor, N. A. Spaldin, D. G. Schlom, R. Ramesh, Large resistivity modulation in mixed-phase metallic systems, Nat. Commun. 6 (2015) 5959.

43. W. Witczak-Krempa, G. Chen, Y. B. Kim, L. Balents, Correlated quantum phenomena in the strong spin-orbit regime, Annu. Rev. Condens. Matter Phys. 5 (2014) 57.

44. X. Huang, L. Zhao, Y. Long, P. Wang, D. Chen, Z. Yang, H. Liang, M. Xue, H. Weng, Z. Fang, X. Dai, G. Chen, Observation of the chiral-anomaly-induced negative magnetoresistance in 3D Weyl semimetal TaAs, Phys. Rev. X 5 (2015) 031023.

45. Z. Q. Liu, J. H. Liu, M. D. Biegalski, J. M. Hu, S. L. Shang, Y. Ji, J. M. Wang, S. L. Hsu, A. T. Wong, M. J. Cordill, B. Gludovatz, C. Marker, H. Yan, Z. X. Feng, L. You, M. W. Lin, T. Z. Ward, Z. K. Liu, C. B. Jiang, L. Q. Chen, R. O. Ritchie, H. M. Christen, R. Ramesh, Electrically reversible cracks in an intermetallic film controlled by an electric field, Nat. Commun. 9 (2018) 41.

46. Z. Q. Liu, C. J. Li, W. M. Lü, X. H. Huang, Z. Huang, S. W. Zeng, X. P. Qiu, L. S. Huang, A. Annadi, J. S. Chen, J. M. D. Coey, T. Venkatesan, Ariando, Origin of the Two-Dimensional Electron Gas at $LaAlO_3$/$SrTiO_3$ Interfaces: The role of oxygen vacancies and electronic reconstruction, Phys. Rev. X 3 (2013) 021010.

47. Z. Q. Liu, L. Li, Z. Gai, J. D. Clarkson, S. L. Hsu, A. T. Wong, L. S. Fan, M. W. Lin, C. M. Rouleau, T. Z. Ward, H. N. Lee, A. S. Sefat, H. M. Christen, R. Ramesh, Full electroresistance modulation in a mixed-phase metallic alloy, Phys. Rev. Lett. 116 (2016) 097203.

48. I. J. Park, T. Lee, P. Das, B. Debnath, G. P. Carman, R. K. Lake, Strain control of the Néel vector in Mn-based antiferromagnets, Appl. Phys. Lett. 114 (2019) 142403.

49. P. Lukashev, R. F. Sabirianov, K. Belashchenko, Theory of the piezomagnetic effect in Mn-based antiperovskites, Phys. Rev. B 78 (2008) 184414.

50. Q. Qin, L. Liu, W. N. Lin, X. Y. Shu, Q. D. Xie, Z. Lim, C. J. Li, S. K. He, G. M. Chow, J. S. Chen, Emergence of topological Hall effect in a SrRuO$_3$ single layer, Adv. Mater. 31 (2019) 1807008.

51. M. Ziese, L. Jin, I. Lindfors-Vrejoiu, Unconventional anomalous Hall effect driven by oxygen-octahedra-tailoring of the SrRuO$_3$ structure, J. Phys.: Mater. 2 (2019) 034008.

52. Y. D. Gu, Y. W. Wei, K. Xu, H. R. Zhang, F. Wang, F. Li, M. S. Saleem, C. Z. Chang, J. R. Sun, C. Song, J. Feng, X. Y. Zhong, W. Liu, Z. D. Zhang, J. Zhu, F. Pan, Interfacial oxygen-octahedral-tilting-driven electrically tunable topological Hall effect in ultrathin SrRuO$_3$ films, J. Phys. D: Appl. Phys. 52 (2019) 404001.

53. Y. Zhang, J. Zelezny, Y. Sun, J. van den Brink, B. H. Yan, Spin Hall effect emerging from a noncollinear magnetic lattice without spin-orbit coupling, New J. Phys. 20 (2018) 073028.

54. X. R. Wang, C. J. Li, W. M. Lü, T. R. Paudel, D. P. Leusink, M. Hoek, N. Poccia, A. Vailionis, T. Venkatesan, J. M. D. Coey, E. Y. Tsymbal, Ariando, H. Hilgenkamp, Imaging and control of ferromagnetism in LaMnO$_3$/SrTiO$_3$ heterostructures, Science 349 (2015) 716.

55. L. M. Zheng, X. R. Wang, W. W. Lü, C. J. Li, T. R. Paudel, Z. Q. Liu, Z. Huang, S. W. Zeng, K. Han, Z. H. Chen, X. P. Qiu, M. S. Li, S. Z. Yang, B. Yang, M. F. Chisholm, L. W. Martin, S. J. Pennycook, E. Y. Tsymbal, J. M. D. Coey, W. W. Cao, Ambipolar ferromagnetism by electrostatic doping of a manganite, Nat. Commun. 9 (2018) 1897.

56. S. J. May, J. W. Kim, J. M. Rondinelli, E. Karapetrova, N. A. Spaldin, A. Bhattacharya, P. J. Ryan, Phys. Rev. B 82 (2010) 014110.

57. L. Qiao, J. H. Jang, D. J. Singh, Z. Gai, H. Y. Xiao, A. Mehta, R. K. Vasudevan, A. Tselev, Z. X. Feng, H. Zhou, S. Li, W. Prellier, X. T. Zu, Z. J. Liu, A. Borisevich, A. P. Baddorf, M. D. Biegalski, Dimensionality controlled octahedral symmetry-mismatch and functionalities in epitaxial LaCoO$_3$/SrTiO$_3$ heterostructures, Nano Lett. 15 (2015) 4677.

**Figure 1**

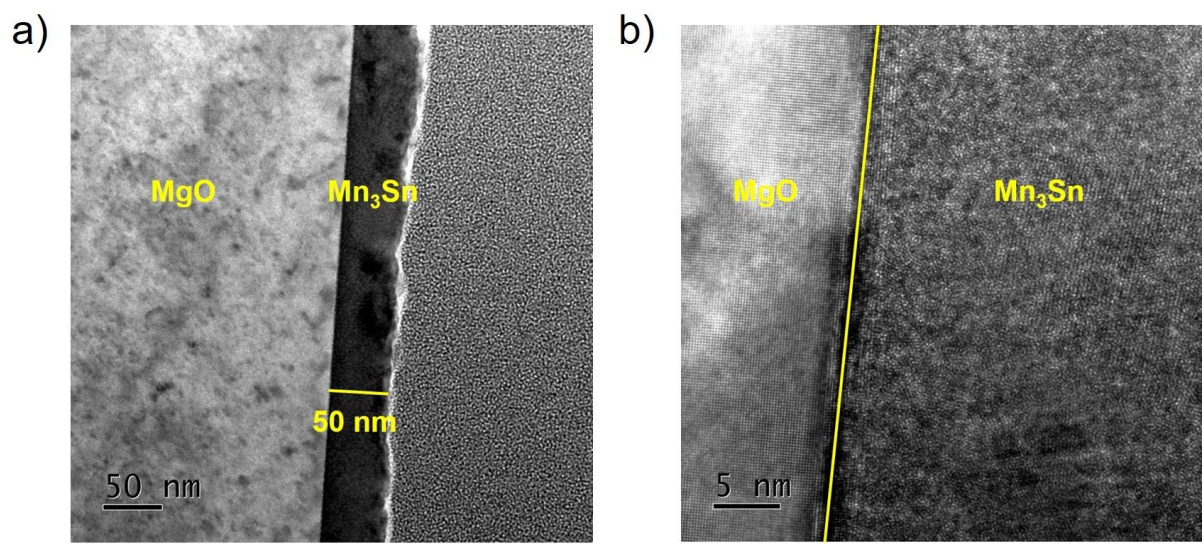

**Fig. 1. Transmission electron microscopy images.** a) Cross-section image of a 50-nm-thick $Mn_3Sn$/MgO heterostructure. b) Zoom-in image of an interfacial region. The scale bar is 5 nm.

**Figure 2**

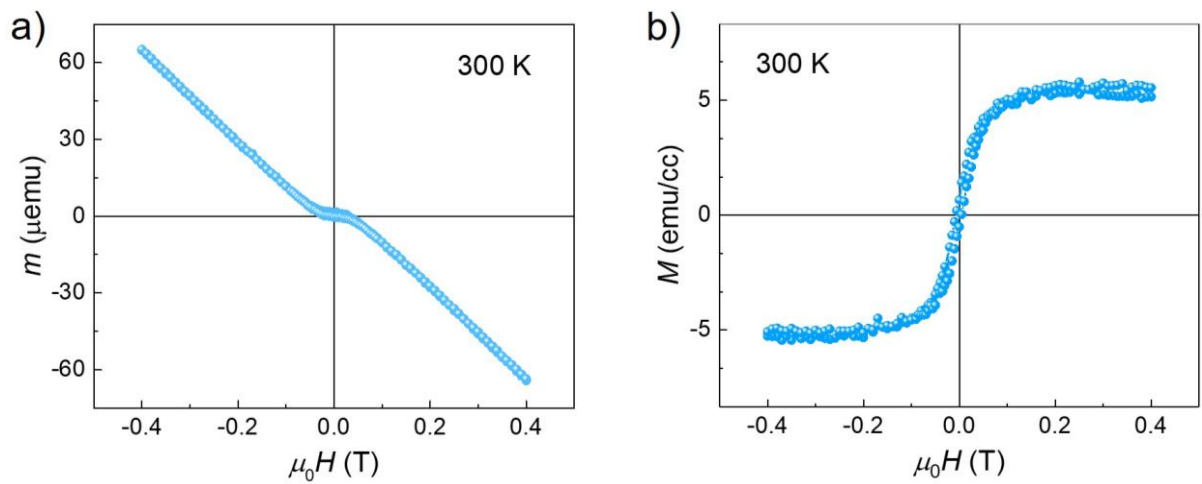

**Fig. 2. Mangetism of Mn₃Sn in the Mn₃Sn/MgO hetreostructure.** a) Magnetic-field dependent magnetic moment of the Mn$_3$Sn/MgO heterostructure at 300 K. b) Extracted ferromagnetic moment of the Mn$_3$Sn film versus magnetic field after the subtration the diamagentic background of the MgO substrate.

# Figure 3

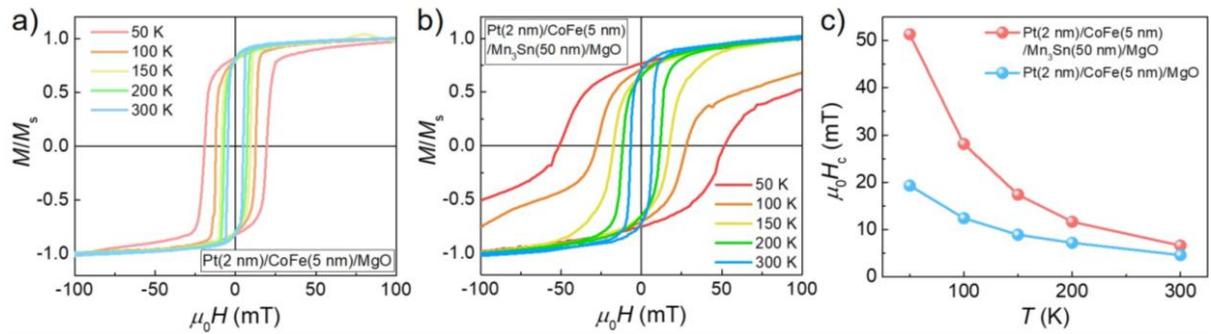

**Fig. 3. Exchange coupling.** a) Magnetization versus magentic field (*M-H*) of a 5-nm-thick $Co_{90}Fe_{10}$ (CoFe) layer grown on MgO capped by a 2-nm-thick Pt layer at different temperature. b) *M-H* loops of a 5-nm-thick CoFe layer grown on a 50-nm-thick $Mn_3Sn$ thin film. c) Coercicivty field ($\mu_0 H_c$) of CoFe in the two types of heterostructures extracted from a & b.

**Figure 4**

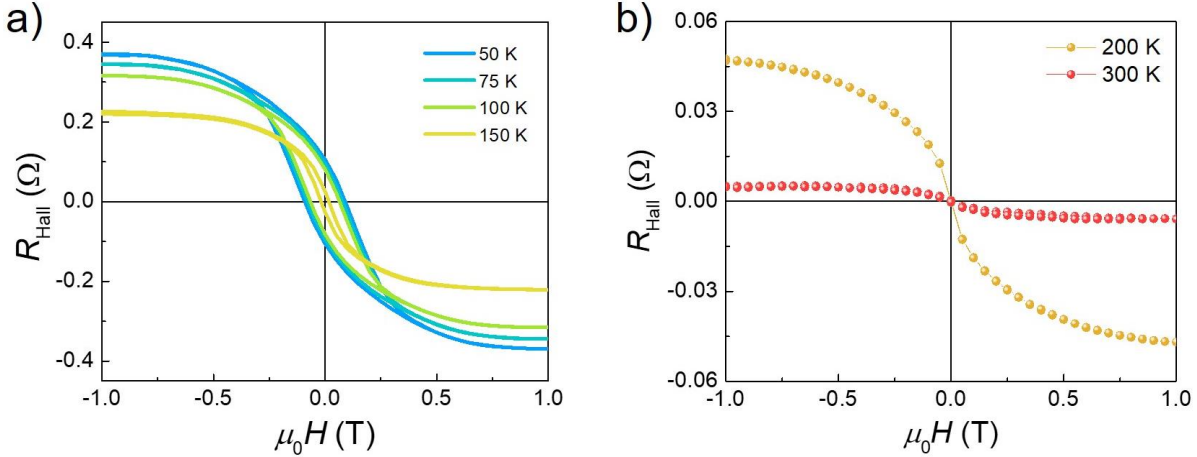

**Fig. 4. Anomous Hall effect (AHE) in the Mn₃Sn/MgO heterostructure.** a) AHE from 50 to 150 K measured up 1 T. b) AHE at 200 and 300 K measured up 1 T.

**Figure 5**

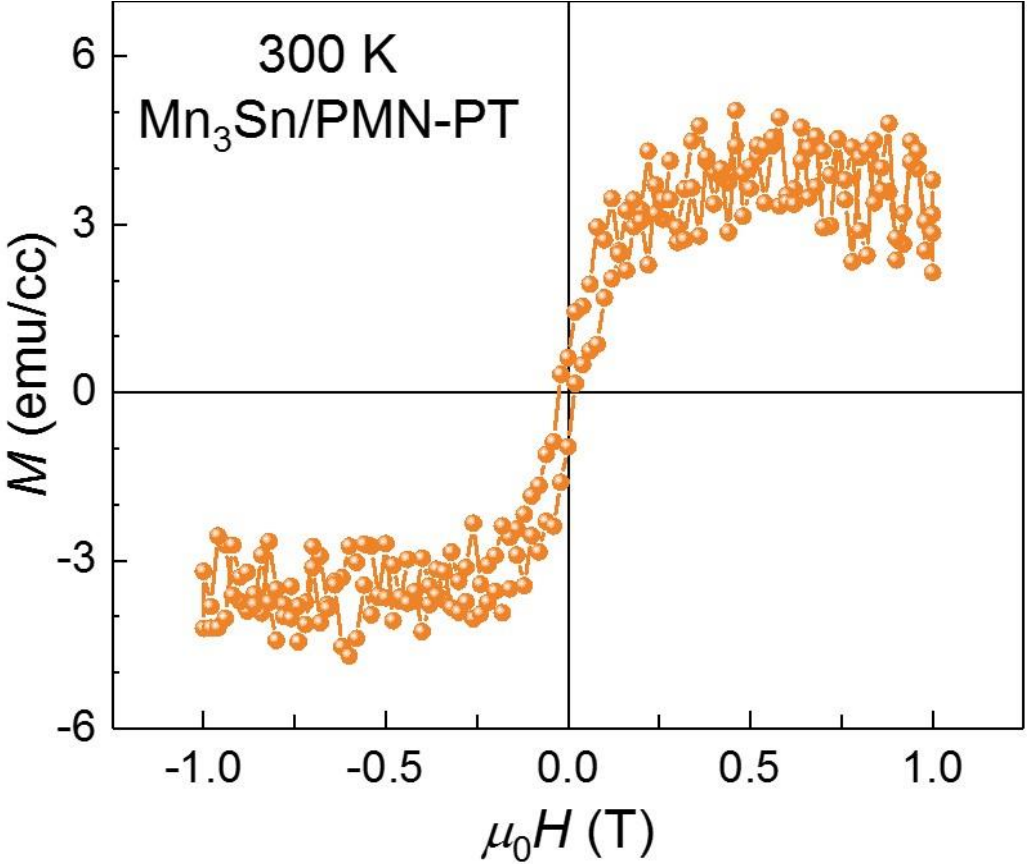

**Fig. 5.** Mangetism of the Mn$_3$Sn film in a Mn$_3$Sn/PMN-PT hetreostructure measured up to 1 T at 300 K.

**Figure 6**

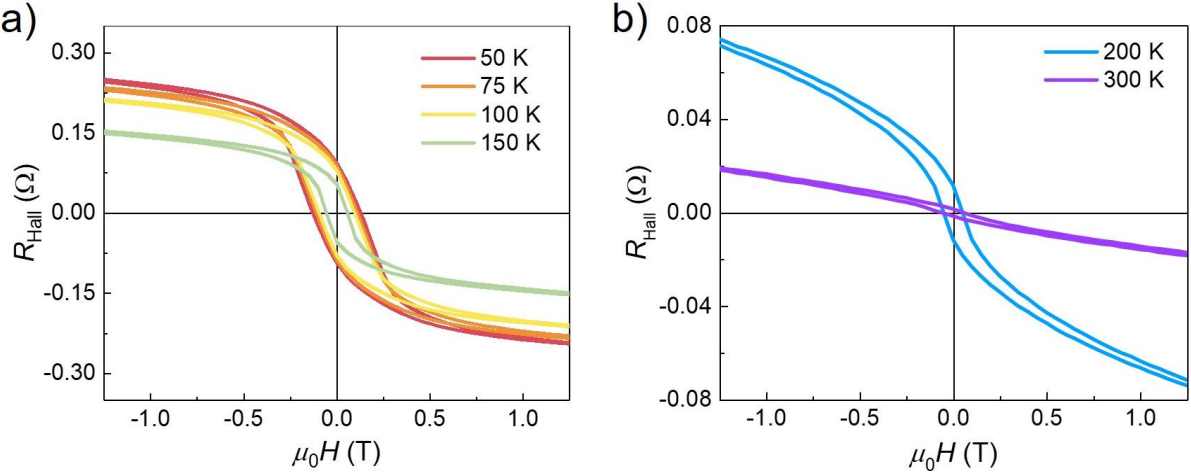

**Fig. 6. Anomous Hall effect (AHE) in the Mn$_3$Sn/PMN-PT heterostructure.** a) AHE from 50 to 150 K. b) AHE at 200 and 300 K.

# Figure 7

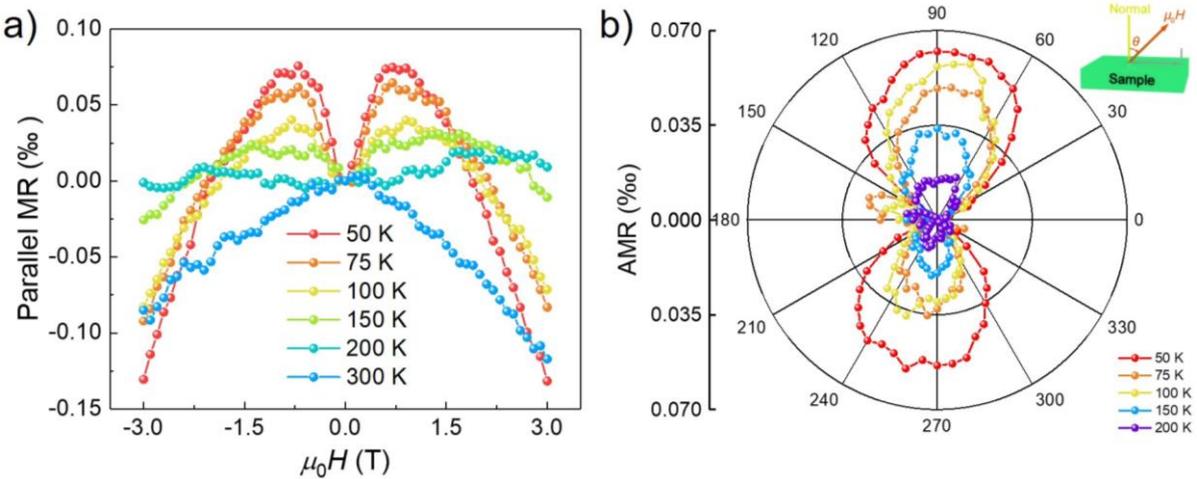

**Fig. 7. Magnetoresistance.** a) Magnetoresistance at various temperatures while the magnetic field is parallel to the measuring current. b) Polar plot of the anisotropics magnetoresistance at different temperatures. Inset: Schematic of the measurement geometry.

# Figure 8

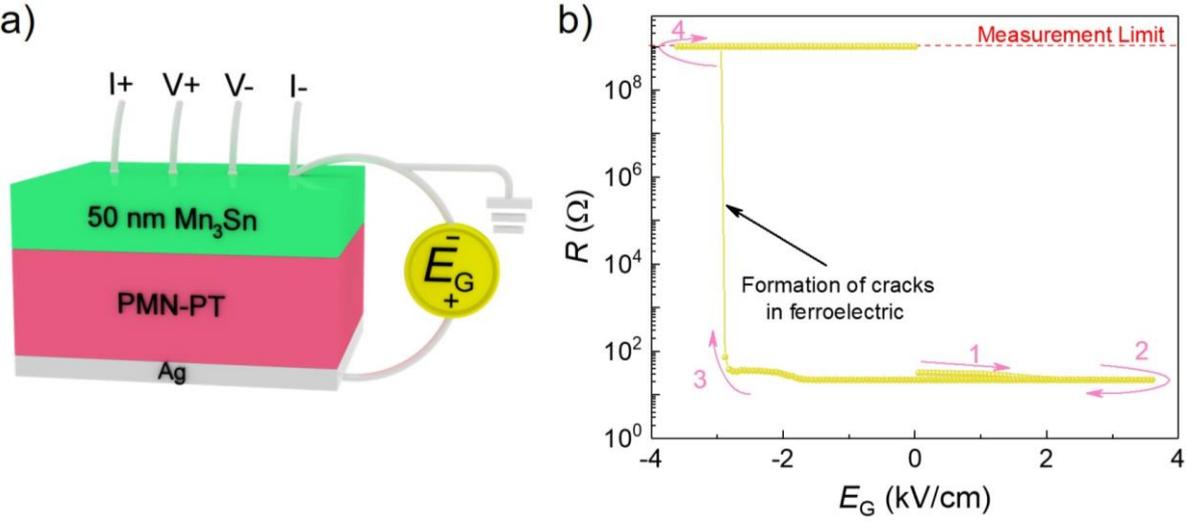

**Fig. 8. Formation of cracks in PMN-PT at room temperature.** a) Sketch of the gate-field ($E_G$) controlled resistance measurement geometry. b) Resistance versus $E_G$.

# Figure 9

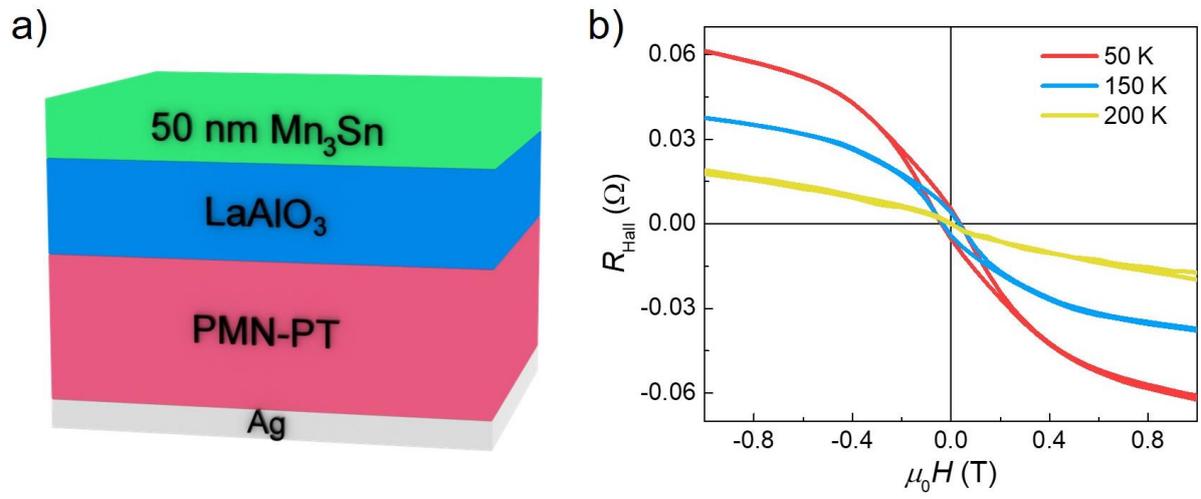

**Fig. 9. AHE in a Mn₃Sn/LaAlO₃/PMN-PT heterostructure.** a) Schematic of multilayer heterostructure. b) AHE at 50, 150 and 200 K.

# Figure 10

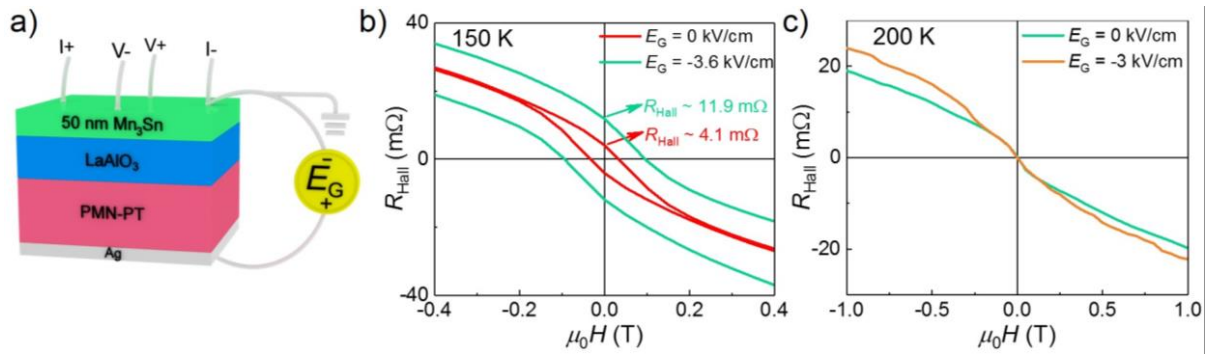

**Fig. 10. Proof-of-concept of an electric-field controlled topological antiferromagnetic spintronic device.** a) Schematic of $E_G$-controlled AHE. b) AHE at 150 K under $E_G$ = 0 and -3.6 kV/cm. **c,** AHE at 200 K under $E_G$ = 0 and -3 kV/cm. The non-zero electric fields were applied at room temperature and then kept onto the multiferroic heterostructure during cooling down to 150 and 200 K.